\pdfoutput=1

\documentclass[11pt]{article}

\usepackage[]{acl}
\usepackage{xcolor,colortbl}
\usepackage{todonotes}
\usepackage{times}
\usepackage{latexsym}
\usepackage{soul}
\usepackage[T1]{fontenc}

\usepackage[utf8]{inputenc}

\usepackage{microtype}
\usepackage{graphicx}
\usepackage{tabularx}
\usepackage{makecell}
\usepackage{multirow}
\usepackage{subcaption}
\usepackage{enumitem}

\newcommand{\franitem}{\item}

\title{User-Driven Research of Medical Note Generation Software}

\author{Tom Knoll\textsuperscript{1}, Francesco Moramarco\textsuperscript{1,2}, Alex Papadopoulos Korfiatis\textsuperscript{1}, \\
\bf{Rachel Young\textsuperscript{1}}, \bf{Claudia Ruffini\textsuperscript{1}},
\bf{Mark Perera\textsuperscript{1}},
\bf{Christian Perstl\textsuperscript{1}},\\
\bf{Ehud Reiter\textsuperscript{2}},
\bf{Anya Belz\textsuperscript{2,3}},
\bf{Aleksandar Savkov\textsuperscript{1}}
\\
\textsuperscript{1}Babylon
\textsuperscript{2}University of Aberdeen
\textsuperscript{3}ADAPT Research Centre, Dublin City University
\\
\textsuperscript{1} \texttt{\{tom.knoll, francesco.moramarco, alex.papadopoulos,}
\\ 
\texttt{rachel.young, claudia.ruffini, mark.perera,}
\\
\texttt{christian.perstl, sasho.savkov\}@babylonhealth.co.uk}
\\
\textsuperscript{2} \texttt{\{r01fm20, ehud.reiter, anya.belz\}@abdn.ac.uk}}

\begin{document}
\maketitle

\begin{abstract}

A growing body of work uses Natural Language Processing (NLP) methods to automatically generate medical notes from audio recordings of doctor-patient consultations. However, there are very few studies on how such systems could be used in clinical practice, how clinicians would adjust to using them, or how system design should be influenced by such considerations. In this paper, we present three rounds of user studies, carried out in the context of developing a medical note generation system. We present, analyse and discuss the participating clinicians' impressions and views of how the system ought to be adapted to be of value to them. Next, we describe a three-week test run of the system in a live telehealth clinical practice. Major findings include (i) the emergence of five different note-taking behaviours; (ii) the importance of the system generating notes in real time during the consultation; and (iii) the identification of a number of clinical use cases that could prove challenging for automatic note generation systems.

\end{abstract}

\section{Introduction}
\label{sec:introduction}

\begin{figure}[t]
    \centering
    \includegraphics[width=.5\textwidth]{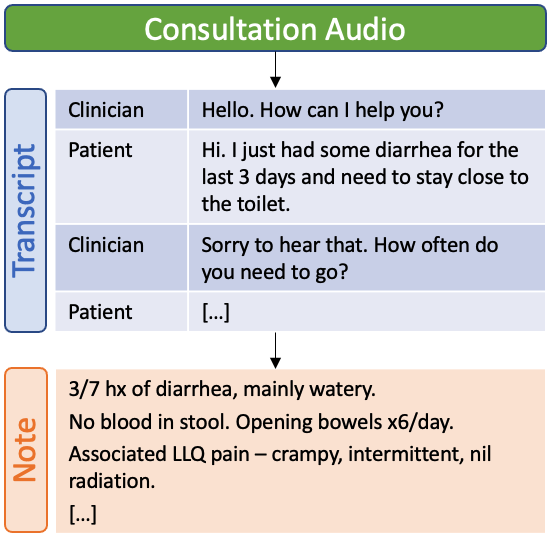}
    \caption{A typical Note Generation pipeline, where the audio recording of the medical consultation is automatically transcribed with Speech-to-text software and then summarised into consultation notes with sequence-to-sequence models.}
    \label{fig:notegen}
\end{figure}

With the introduction of Electronic Health Records (EHR), clinicians are required to keep a detailed record of each patient interaction \cite{menachemi2006reviewing}. While this creates a wealth of useful data and may lead to better medical outcomes, \citet{arndt2017tethered} show that the burden of administrative tasks is a major contributor to clinician burnout.
To address this, some recent studies \cite{zhang2018learning, enarvi2020generating, joshi2020dr, zhang2021leveraging} propose to use Speech Recognition to transcribe the audio of a medical consultation and then to train sequence-to-sequence models to summarise the transcript into a consultation note (Figure \ref{fig:notegen}). While intrinsic evaluations\footnote{Evaluations that involve subjects being shown both NLG and human-written texts and being asked to compare their ratings \cite{reiter-belz-2009-investigation}.} \cite{moramarco2021preliminary, yim2021towards, chintagunta2021medically, moramarco2022human} show that these methods may be effective at capturing the salient points of a medical consultation, most studies focus on the technical difficulties of developing the systems with little consideration for the Human-Computer Interaction (HCI) and usability challenges involved in putting such a system into clinical practice. This is especially important for ``human in the loop'' systems, like Note Generation, that still involve manual checking of any automatically generated content.

In this paper we present methodology and findings of three rounds of user research and design: (i) \emph{Current Note-Taking Discovery} to gather user requirements and initial impressions; (ii) \emph{Initial User Interface (UI) Testing} where clinicians were shown video mock-ups of three potential design options and asked to provide feedback; and (iii) \emph{Mock Consultations}, in which the clinicians interacted with a Wizard-of-Oz \cite{dahlback1993wizard} prototype, with another clinician acting as the automated system.

Following these, we developed a Note Generation system informed by the user studies and carried out a live test with five clinicians who used the system to aid them in the task of writing a consultation note for each consultation. The experiment lasted three weeks, during which over 300 consultations were carried out with the system.

We document the insights gathered through interviews, post-consultation surveys, and analysis of the generated notes, reporting the major requirements for developing a Note Generation system into a useful product while highlighting the importance of user research and design alongside NLP research. Among our main findings are the emergence of five note-taking behaviours (Figure \ref{fig:note_taking_styles}) and the need for the Note Generation system to produce output in real time during the consultation. We also identify three clinical use cases that could prove challenging to a Note Generation system: patient dishonesty, multiple presenting complaints, and clinicians' non-verbal observations.

\section{Related Work}
\label{sec:related_work}

The task of automatically generating consultation notes from clinician-patient interactions is rapidly gaining interest in the research community \cite{zhang2018learning, macavaney2019ontology, enarvi2020generating, joshi2020dr, chintagunta2021medically, yim2021towards, moramarco2022human, zhang2021leveraging}. The prevalent methods employ pre-trained, generic summarisation models and fine-tune them on dedicated datasets of consultation transcripts and corresponding notes. However, the two tasks have differences; in particular, Note Generation has a component of language translation as well as summarisation, whereby the note uses medical terminology and inferred statements that aren't present in the original transcript (Figure \ref{fig:notegen} and Table \ref{tab:transcript-note} for examples). Therefore, generic summarisation evaluation methods and UX principles may not apply to this task.

Some of the Note Generation studies include an intrinsic human evaluation of the generated consultation notes to measure common Natural Language Generation (NLG) criteria such as accuracy, fluency, and completeness; also, \citet{moramarco2021preliminary} propose post-editing as a way clinicians may incorporate the generated notes into their workflows. However, there is very limited work on user needs, design decisions, and system requirements for a note-taking tool to be useful in live clinical practice. In this paper, we aim to address these questions for the task of medical note generation. 

In the broader medical domain, there has been a marked increase in studies on the interaction between human users and different tools with clinical applications since 2013 \cite{stowers2018human}, with usability-related studies making up 25\% of them.
\citet{clarke2013addressing} review 17 papers that identify human-computer interaction (HCI) issues in the use of Electronic Health Records in medical practice, and categorise them into four types: poor display of information, cognitive overload, navigation issues and workflow issues. 
\citet{info:doi/10.2196/humanfactors.9481} investigate the UX experience of patients managing their type 2 diabetes with an in-home monitoring device. The authors conduct a contextual inquiry followed by semi-structured interviews with 9 patients and report the patients' experiences and emotions while using the device, perceived benefits and limitations.
\citet{MEGGES2018636} present a usability study of wearable GPS devices for people with dementia. The study ran for 4 weeks and comprised two alternative devices and 17 participants. It measured clinical effectiveness, revealed a clear preference for one of the two devices, and informed subsequent design decisions (such as software features, buttons, and battery life).

In the NLG space, \citet{REITER200341} conduct a clinical trial with 2553 smokers to show whether system-generated tailored cessation letters would be more effective than a generic cessation letter written by medical practitioners.
\citet{portet2009automatic} carry out a clinical off-ward experiment in a Neonatal Intensive Care Unit where medical practitioners compare the effect on decision-making that a data-to-text system has versus human expert textual descriptions of neonatal signals.
\citet{10.1145/1518701.1518891} describe an animated, empathic virtual nurse interface for counseling hospital patients with low health literacy. Their results indicate patients found the system easy to use, were satisfied with the system, and most said they preferred receiving the discharge information from the agent over their doctor or nurse.

\section{UX Exploration}
\label{sec:discovery}

This work was conducted at Babylon, a digital healthcare company that employs clinicians to carry out virtual, video consultations with patients through a platform referred to as the Clinical Portal. As part of their workflow, Babylon clinicians write medical notes for each consultation; the notes are useful for any future healthcare professionals interacting with the patient, but they are also made available to the patient after the consultation and serve as a medico-legal record of the interaction.

In order to investigate the value of a Note Generation system and to design an interface that best supports clinicians, we carried out three experiments with the following aims: (i) learning about the clinicians' current note-taking behaviours (Section \ref{subsec:phase1}), (ii) gathering their initial thoughts about Note Generation systems using low-fidelity designs (Section \ref{subsec:phase2}), and (iii) collecting hands-on insights from using fully interactive designs (Section \ref{subsec:phase3}).

The research participants throughout the study were all UK clinicians with diverse backgrounds and at least one year's experience working at Babylon, who regularly used the Clinical Portal to carry out consultations with patients. As part of our ethical consideration, we obtained consent from the participants to use their feedback for research and development (excluding marketing purposes). Their time was paid at £70 per hour\footnote{A rate higher than their standard pay, in order to discourage cancellations.} and they were aware they could withdraw at any point.

\subsection{Phase One: Current Note-Taking Discovery \& Initial Reactions to Note Generation}
\label{subsec:phase1}

Prior to the creation of any interface designs for the Note Generation system, we conducted a round of user research to get a detailed understanding of how the clinicians currently record notes during consultations, their thoughts about the tools they use, factors that impede effective note-taking, and initial impressions on using Note Generation systems to support note-taking.

\begin{figure*}[t]
    \centering
    \includegraphics[width=1.0\textwidth]{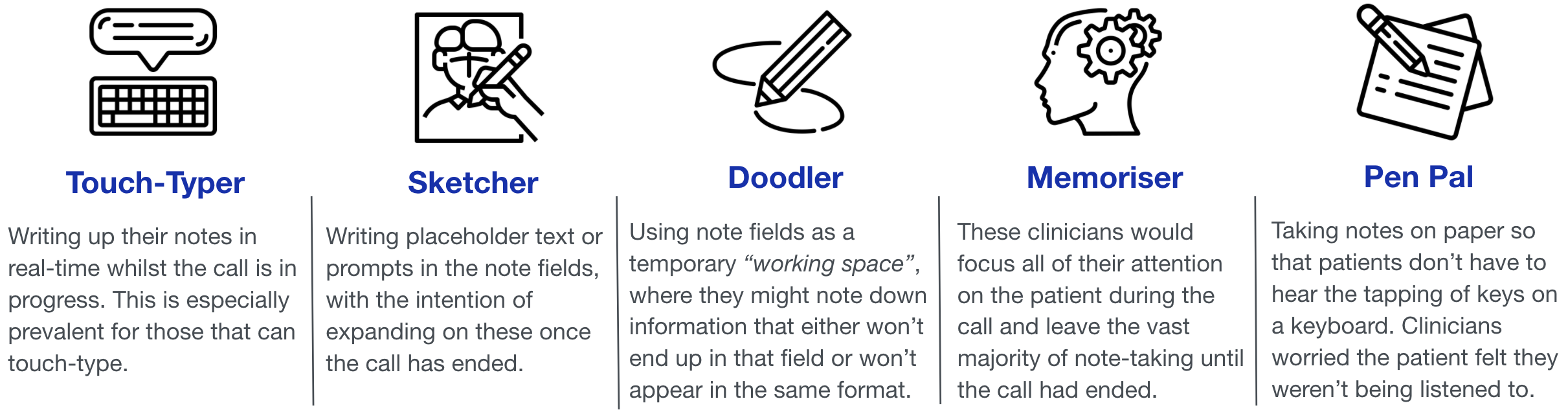}
    \caption{The five note-taking personas emerged from the Discovery round of user research.}
    \label{fig:note_taking_styles}
\end{figure*}

\subsubsection{Methods}
\label{subsubsec:phase1-methods}

Seven UK clinicians (four women and three men) across multiple clinical disciplines, including five UK General Practitioners (GPs) and two Advanced Nurse Practitioners, participated in 60-minute one-on-one interviews with a User Researcher (the lead author). The focus of these sessions was to understand current note-taking behaviours. Discussion topics included:

\begin{itemize}[leftmargin=.4cm]
    \itemsep-.2em
    \franitem What constitutes a “good note”
    \franitem The format and structure of consultation notes
    \franitem Balancing note-taking and interacting with patients
    \franitem Time management and note-taking
    \franitem Reviewing notes prior to submission
    \franitem Role of note-taking in decision making
    \franitem Initial thoughts about supporting note-taking with automated systems
\end{itemize}

The study was carried out through semi-structured interviews, with a list of pre-determined open questions we asked all participants. One or two note-takers were present in each interview, transcribing everything the participant said. The video and audio of the interviews were also captured (for reference should the notes not be clear or detailed enough). All qualitative feedback was then annotated with a standard UX software,  Dovetail.\footnote{\url{https://dovetailapp.com/}}

\subsubsection{Findings}

The research sessions revealed several insights that influence the 
way consultation notes are captured:

\begin{itemize}[leftmargin=.4cm]
    \itemsep-.2em
    \franitem Six of the seven participants said that having limited consultation time (typically ten minutes in the UK) forces them to capture notes while talking to the patient, rather than waiting until after the call to write up their notes. This is because delaying note-taking until after the call would make them late for their next appointment. They reported they would prefer to dedicate their full attention to the patient during the call, rather than having to capture notes simultaneously.

    \franitem In order to minimise the amount of manual typing required to write up consultation notes, two clinicians reported using text expanders \cite{LACKEY2014481} (macros that trigger a pre-written section of text when a keyword is typed), three use pre-written templates (such as Word documents, created by clinicians themselves, that they copy and paste from), and one clinician uses dictation software to speed up the creation of action plans.

    \franitem Two out of the seven clinicians find value in the manual act of note-taking to help them organise their thought processes and guide them towards a suitable action plan. However, the remaining five clinicians saw note-taking as a record-keeping exercise rather than something that improves their medical decision making. To these clinicians, having to type up notes is an administrative task that takes up time they would rather be spending with patients.

    \franitem All participating clinicians reported that they review  the notes that they have taken before finalising a consultation (which is most often after the end of the call with the patient). Common edits during this review include resolving spelling and grammatical errors, restructuring some of the notes and expanding any placeholder text they might have added.
    
    \franitem Finally, clinicians identified some areas that could be hard to handle for an automatic Note Generation system:

    \begin{itemize}[leftmargin=.4cm]

        \item \textbf{Patient dishonesty.} In some cases, patients could withhold information or exaggerate their symptoms \citep{palmieri_lies_2009}. According to six of the seven participants, recording notes in situations where they believe a patient is being dishonest is challenging, as Babylon patients have access to the consultation notes. Clinicians need to be diplomatic and avoid conflict while still making sure that the notes accurately reflect their own justification for diagnosis and treatment decisions.
    
        \item \textbf{Multiple presenting complaints.} Sometimes, patients seek advice for multiple medical issues. Two clinicians mentioned trying to dissuade patients from doing this. However, all clinicians reported they would attempt to address additional patient problems if time allows. Five of them explained they would deal with this by separating the notes for different medical issues into separate paragraphs.
        
        \item \textbf{Non-verbal communication.} All clinicians mentioned their notes often include information that is not verbalised, such as observations from that patient's video feed or their general demeanour; such information would be impossible for a Note Generation system to capture.
    \end{itemize}

\end{itemize}

The research also revealed five note-taking behaviours that were adapted into mini personas (Figure \ref{fig:note_taking_styles}). These personas have persisted throughout the research studies, with all observed clinicians fitting comfortably into one or more of the five categories.
In all subsequent studies described in this paper, approximately 90\% of interviewed clinicians fall under the ``Touch-Typer'', ``Sketcher'', or ``Doodler'' personas. This provided focus for the types of note-taking behaviours that a Note Generation system could help to support.


\subsection{Phase Two: Initial UI Testing}
\label{subsec:phase2}


With a better understanding of current note-taking behaviours, an initial low-fidelity design for the system interface was created to gather clinician feedback. Due to the complexity of actually implementing a Note Generation system in real clinical practice, a video mock-up of the design was shown to clinicians.

\subsubsection{Methods}

Five Babylon clinicians (all GPs, three women and two men) who were uninvolved in the previous round of research participated in 60 minute one-on-one semi-structured interviews (following the study design described in Section \ref{subsubsec:phase1-methods}). They were asked to provide feedback on the general concept of Note Generation and on three potential design directions for the Note Generation system. The three design variations (Figure \ref{fig:ui-designs} in the Appendix) simulated how Note Generation content could be made available to them during consultations and were presented to the clinicians as videos of mock consultations. The variations were:

\begin{itemize}[leftmargin=.4cm]
    \itemsep-.2em
    \franitem \textbf{Design 1:} A full live transcript of the conversation between the patient and the clinician being shown in near real time. At the end of the call, the Note Generation system would produce a note based on the entire transcript, and the note would additionally appear on screen (Figure \ref{fig:sub-first}).

    \franitem \textbf{Design 2:} No real time output during the call. At the end of the call, the Note Generation system would produce a note based on the entire transcript, and only that note would appear on screen (Figure \ref{fig:sub-second}).

    \franitem \textbf{Design 3:} The Note Generation system would produce a note and show it on screen during the call in near real time using the transcript available up to that point (Figure \ref{fig:sub-third}).
\end{itemize}

\begin{figure*}[t]
    \centering
    \includegraphics[width=1.0\textwidth]{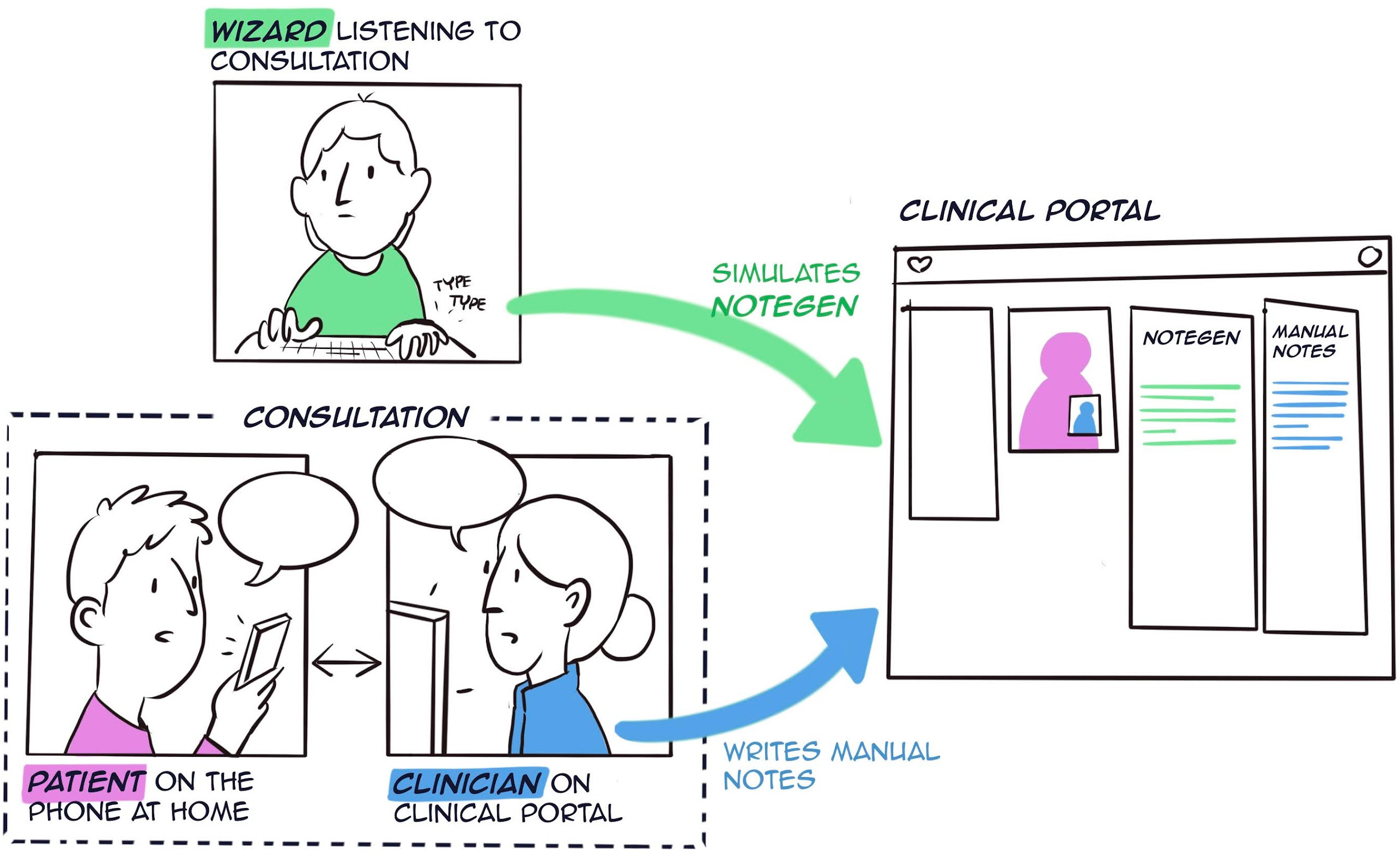}
    \caption{A diagram of the `Wizard of Oz' setup. The clinician participating in the research is having a consultation with a mock patient, and can supplement their own notes with automated ones, generated not by a Note Generation system (NoteGen) but by a `Wizard' listening to the conversation.}
    \label{fig:wizard-of-oz}
\end{figure*}

\subsubsection{Findings}

Overall, all clinicians found the concept of the Note Generation system useful and believed that a tool that could produce consultation notes would help them to focus more on their patients due to the reduced need for them to type simultaneously.
However, they also raised concerns about perceived limitations of a Note Generation system, including the worry that it would take control of the notes away from them; that it may just be a straight transcription service; and that it may fail to capture notes with medical terminology and instead report layperson's terminology (Table \ref{tab:transcript-note} in the Appendix shows the stylistic differences of transcripts and notes).

Contrary to our initial belief that the main value of a Note Generation system would be time-saving, all clinicians felt that the tool would be more useful in improving the quality of the consultation by allowing them to focus more on the patient.

When shown the potential designs, all clinicians reacted negatively to the verbatim transcript being shown on their screen (Design 1), as it would be too distracting due to the large amount of text and information and the inherent delay of displaying each sentence. Furthermore, all clinicians expressed concerns at having to wait until the end of the consultation for the generated note to appear (Designs 1 and 2) as they wouldn't know whether the model was capturing all the salient points. The real time note display of Design 3 was strongly preferred, as it would allow them to easily check whether the model is correct during the consultation; the real time note could also act as a cognitive artefact, reminding them what they've already asked.
This finding indicates that the output speed of the Note Generation tool might impact clinician uptake; if too slow, clinicians would need to revert back to manual note-taking to avoid missing important patient information.

\begin{figure*}[t]
    \centering
    \includegraphics[width=1.0\textwidth]{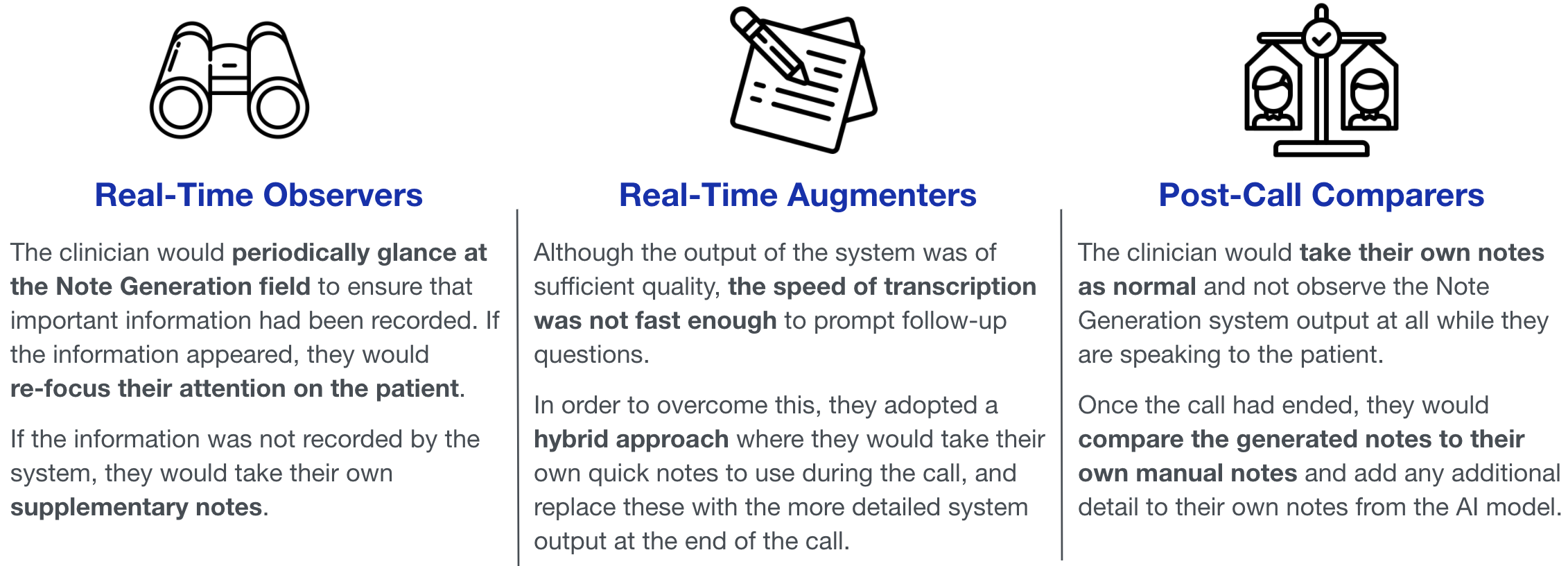}
    \caption{The three styles of interacting with the system that emerged from the Mock Consultation study.}
    \label{fig:behaviours}
\end{figure*}


\subsection{Phase Three: Mock Consultations using Wizard of Oz}
\label{subsec:phase3}

Up to this point, Babylon clinicians had been providing feedback on designs without being able to interact with the system in any way.
In order to allow clinicians to experience what using a Note Generation system might actually be like, we used a Wizard-of-Oz approach. This was an inexpensive way to allow hands-on interaction as well as a better replication of the complex nature of real-world consultations.

\subsubsection{Methods}

Five Babylon GPs (four women, one man) who were uninvolved with any of the previous rounds of research were recruited for 90 minute one-on-one workshops. They were asked to conduct mock consultations with a remote patient, using a mocked-up representation of the Clinical Portal, with a prototype version of the Note Generation system available to them to support their note-taking. We then conducted semi-structured interviews to gather their feedback (following the study design described in Section \ref{subsubsec:phase1-methods}).

The Clinical Portal mock-up, with an additional section for displaying the generated notes, was recreated in Google Slides\footnote{\url{https://www.google.co.uk/slides/about/}}, which was chosen as a freely available application allowing multiple users to edit a document simultaneously. The clinician (research participant) was able to record their own manual notes as well as having the option to supplement these with the Note Generation output. Behind the scenes, the generated notes were not produced by an automated system, but by another Babylon clinician, the `Wizard' (one of the authors), who had been instructed to respond as the Note Generation system would: to insert the notes (one sentence at a time) in the mock Clinical Portal by editing the appropriate field.

In order to make the job of the Wizard easier, the patient (an actor\footnote{British English speaker, paid £30 per hour.}) was asked to present with one of a series of predefined medical complaints that were tightly scripted on case cards; this allowed the Wizard to use pre-written responses that matched with the script being followed by the patient.

In total, the five clinicians (research participants) were each able to conduct three mock consultations over video call which spanned common primary medicine topics: urinary tract infection, sinusitis, and mental health. A diagram of the above setup can be found in Figure \ref{fig:wizard-of-oz}.

\subsubsection{Findings}
The interactive research sessions revealed a number of Note Generation usage patterns:

\begin{itemize}[leftmargin=.4cm]
    \itemsep-.2em
    \franitem We observed that all clinicians' first consultations were more disjointed, as they were glancing at the system output from time to time instead of focusing on the patient. In subsequent mock consultations, they took fewer manual notes and relied more on the system. By the end of the research session all clinicians were observed to use parts of the automated notes in their own note-taking.
    \franitem After using the system, two clinicians reported that the generated note output speed was too slow. When asked, all clinicians
    mentioned that a close to real time speed would be highly desirable, as it would allow them to confirm that important information has been captured by the Note Generation system before the conversation moves onto another topic.
    
    \franitem Based on the way they interacted with the Note Generation system output, we categorised the participating clinicians in three different styles, described in Figure \ref{fig:behaviours}. All clinicians were `Post-Call Comparers' in their first mock consultation; yet, by their last consultation, 40\% of them used the Real-Time Augmenting style, with the rest divided between `Post-Call Comparers' (35\%) and `Real-Time Observers' (25\%).

    \franitem When prompted, all research participants reported that the Note Generation system output was of value to the note-taking process.
    
    \item Two clinicians were observed reading the generated note back to the patient towards the end of the call as a way of confirming the notes matched the patient's perspective.
    
    \item Clinicians either copied the whole generated note before post-editing it (two thirds of all consultations), or copied selected statements into their own notes (one third of all consultations).
    
\end{itemize}

\section{Note Generation system}

The UX Exploration detailed so far gave us, among other insights, indications that: (i) a Note Generation tool could be of value to clinicians; (ii), it would need to generate notes in real time, in order for clinicians to evaluate its output as the consultation is progressing; and (iii), clinicians would still want to be able to take manual notes during the consultation.

Prior to this UX Exploration, we had developed a Note Generation system based on a publicly available BART sequence-to-sequence model \citep{lewis-etal-2020-bart} pre-trained on the CNN summarisation dataset \cite{nallapati2016abstractive}, which we further fine-tuned on a proprietary Babylon dataset of 10,000 transcripts and consultation notes.
In \citet{moramarco2022human} we detail a non-production evaluation of the system, and in \citet{primock2022} we release the mock transcript and note dataset used to evaluate it. Examples of model inputs and outputs from this mock dataset can be found in Appendix \ref{app:transcript-note}.

The following modifications were made as a result of the three studies in the UX Exploration:
\begin{itemize}
    \item The previous system could only generate notes in an end-to-end fashion from the complete consultation transcript. This is not compatible with the clinicians' preferred design, so we retrained the model to produce notes from incremental chunks of the transcript. This was achieved by mapping sentences of the notes and sentences of the transcript with a similarity function and generating incremental training pairs for each consultation. This system will only append text to the previously generated note, preserving anything that the clinician might have previously validated.
    \item Automatically generated notes often contain inaccuracies or omissions \cite{moramarco2022human}. Using a Note Generation system in a live clinical setting might mislead the clinician into introducing errors in their consultation notes or influence the clinical decision making process. In order to mitigate those risks, we re-trained the Note Generation system to only generate the history and examination portions of the notes \cite{pearce2016essential}, preventing it from producing any diagnosis or treatment advice by removing it from the training data.
\end{itemize}

\section{Live Test}
In order to test the usability of the updated Note Generation system in clinical practice, we incorporated the real-time system into the Clinical Portal
and allowed a limited number of clinicians to use it in a supervised environment for a period of three weeks.
Our goal from this live test was to answer the following questions:

\begin{itemize}[leftmargin=.4cm]
    \itemsep-.3em
    \item How well is the Note Generation system able to cope in a live setting?
    \item How many consultations does it take clinicians to start making use of the output from the Note Generation system?
    \item Are there any safety issues that need to be addressed?
    \item What is the current and potential value of the Note Generation system during consultations?
    \item Are there any usability issues that need to be addressed?
\end{itemize}

\subsection{Methods}
In order to maximise the learnings from the live test, the research approach needed to allow us to observe the Note Generation system being used live, ask follow-up questions and understand the clinicians' overall experience.
The research approach for the live test had the following steps:

\begin{itemize}[leftmargin=.4cm]
    \itemsep-.3em
    \item Interviews with clinicians before and after the three week live test.
    \item Weekly observation of clinicians using the Note Generation system during real-world consultations.
    \item A post-consultation survey on the use and value of the system during that particular consultation.
    \item Ad-hoc questions about the clinician experience so far via private messages.
    \item Analytics on the Clinical Portal (such as UI event timestamps) to quantify Note Generation system usage.
\end{itemize}
	
As the Note Generation software is a clinician-facing tool and falls under Babylon normal service delivery, the patients were not participants of the live test. However, for all consultations observed, explicit consent to do so was obtained from the patients (consent form in Appendix \ref{sec:appendix}), and the observations were completely passive (no interaction) to minimise inconvenience and distress both to patient and clinician.
As an extra mitigation for any generated note inaccuracy, we also clearly instructed clinicians to continue manually recording the key information of the consultation.

Five Babylon clinicians (GPs, three women, two men) were invited to participate in the three-week live test. The inclusion criteria were: (i) UK clinicians having worked at Babylon for at least 6 months, (ii) consulting at Babylon for at least 10 hours a week, (iii) willing to use the Note Generation system during their consultations for the three-week period, (iv) agreeing to have a small number of their consultations with patients observed by the researchers, (v) minimum bandwidth (download 5Mb/s, upload: 3Mb/s; to support the observing researchers), (vi) be willing to participate in all our research activities related to the project (such as the interviews and post-consultation surveys). The participants' time in consultations was under their normal working hours at Babylon. The extra time for interviews and training was paid at the study standard rate.

Prior to the start of the live test, the clinicians were provided with 60 minutes one-on-one up-front training on how to use the tool. The training consisted of showing them videos of the Note Generation system during a mock consultation, warning them that the system is not able to capture non-verbal observations, and that the medico-legal responsibility for the notes still rested with them (the clinicians). They were then provided with access to the Note Generation system and enabled to use it during their real-world consultations. Through initial interviews, we gathered that three of our participants fit in the Sketchers note-taking style; one in Doodlers; and one in Touch Typers (see Figure \ref{fig:note_taking_styles}).
Throughout the experiment, we observed a total of 4 hours of live consultations for each clinician. Finally, we collected their summary feedback through 60-minute, one-on-one closure interviews.

\begin{table}[t]
    \centering
    \begin{tabular}{c|c|c|c}
        \textbf{Participant} & \textbf{\# Cons.} & \textbf{\% copied} & \textbf{Overlap} \\\hline
        A & 101 & 97\% & 50\% \\
        B & 81 & 23\% & 13\% \\
        C & 69 & 42\% & 22\% \\
        D & 18 & 50\% & 34\% \\
        E & 38 & 13\% & 19\% \\
    \end{tabular}
    \caption[]{\textbf{\# Cons.} shows the number of consultations carried out in the live test, \textbf{\% copied} the amount of generated note copied into clinicians' own notes (calculated with UI copy events), and \textbf{Overlap} an indication of how much of the final note comes from the system. This is estimated with ROUGE-L \cite{lin2004rouge} recall\protect\footnotemark, which measures the n-gram overlap between generated and final notes divided by the n-gram count in the final note.}
    \label{tab:overlap-copy-scores}
\end{table}

\subsection{Findings}

Here are the key points uncovered by observing and interviewing the clinicians using the Note Generation system in live practice:

\begin{itemize}[leftmargin=.4cm]
    \itemsep-.3em
    \item In the initial observation sessions, all clinicians took manual notes, ignoring the output of the Note Generation system. However, by the end of the experiment, all clinicians were making use of the generated notes by copying them at least partly into their notes. Copying occurrence varied between clinicians (see Table \ref{tab:overlap-copy-scores}).

    \item During difficult consultations (for example, consultations with multiple interconnected medical issues, over-explaining patients, or mental health cases), we observed all clinicians reverting to manual note-taking instead of using the Note Generation output.
    
    \item
    Analysis of the notes showed that clinicians always made amendments to the Note Generation output before using it. These included: fixing any identified errors, reorganising the output into distinct paragraphs, changing references of days to dates (`on Saturday' -> `14/01/21'), and changing the use of pronouns to match their preferred style.

    \item In order to measure the impact of the Note Generation tool, we asked all clinicians the following questions both before and after their usage of the tool. The answers to these questions were collected through a rating scale (from very difficult `1' to very easy `10'); we report the average for all 5 clinicians.

    \begin{itemize}[leftmargin=.4cm]
        \itemsep0em
        \item \textit{How easy is it to write up your notes within the ten minutes allocated for the consultation?} \textbf{(before: 4.8, after: 5.5)}
        \item \textit{How easy is it to focus on the patient during consultations?} \textbf{(before: 6.2, after: 8.3)}
        \item \textit{How easy is it to multitask during consultations with patients?} \textbf{(before: 5.8, after: 6.8)}
    \end{itemize}
    
    \item Finally, when asked in the post-experiment interviews, all clinicians reported that they would like to continue using the Note Generation tool in their clinical practice.

\end{itemize}

\footnotetext{\url{https://pypi.org/project/rouge-score/}}

\section{Discussion and Conclusion}
In this paper we presented a series of user research studies and described how they influenced the development of a Natural Language Processing system for automatically generating consultation notes.

Some of the main challenges in putting such a system into clinical practice are not due to the system's technology or limitations, but instead to the user experience design choices around it. User research and live testing is essential to ensure the system is actually useful for clinicians. Among the list of tools available, we found that a low-fidelity Wizard-of-Oz prototype was both incredibly valuable to gather hands-on feedback from clinicians and significantly easier to implement than a functional product prototype.

More specifically, the need to generate notes in real time is a requirement that was identified through the UX Exploration and confirmed in both the Wizard-of-Oz and live test phases. To our knowledge, this has not been covered in similar studies and we believe it will be a key challenge for future systems.
Similarly, the user interviews uncovered three clinical use cases that could prove challenging to a Note Generation system: patient dishonesty, multiple presenting complaints, and clinicians' non-verbal observations.

We observed an interest for the tool beyond the live test, even from clinicians who were initially skeptical of automatically generated notes. We also highlight the individual differences in personal note-taking styles, which mean that clinicians use it in different, but still valuable ways.

One limitation of our work so far is the relatively small sample of clinicians (no more than 7 in each round) and the short time frame (3 weeks) in which they used the tool in our live test. The results we present from three weeks of live testing might not be representative of long-term usage of the Note Generation tool. Another limitation is that the design, based on data and processes from the UK, may not transfer easily to other regions without substantial changes.

In conclusion, we find that user experience studies, currently not commonly carried out in NLP research, revealed essential requirements for the design of our Note Generation system. We believe that to design an interactive tool it is crucial to test with active users from the start and hope that this work may help further studies in the field of medical Note Generation.

\section*{Acknowledgements}
We would like to thank all the clinicians who participated in the UX studies and the live test of the Note Generation system. We are grateful to Oriol Valldeperas for the beautiful diagram of the Wizard of Oz setup. Lastly, a warm `thank you' to all the patients who have consented to their consultations being observed for the purpose of this study.

\bibliography{anthology,custom}

\begin{thebibliography}{27}
\expandafter\ifx\csname natexlab\endcsname\relax\def\natexlab#1{#1}\fi

\bibitem[{Arndt et~al.(2017)Arndt, Beasley, Watkinson, Temte, Tuan, Sinsky, and
  Gilchrist}]{arndt2017tethered}
Brian~G Arndt, John~W Beasley, Michelle~D Watkinson, Jonathan~L Temte, Wen-Jan
  Tuan, Christine~A Sinsky, and Valerie~J Gilchrist. 2017.
\newblock Tethered to the ehr: primary care physician workload assessment using
  ehr event log data and time-motion observations.
\newblock \emph{The Annals of Family Medicine}, 15(5):419--426.

\bibitem[{Bickmore et~al.(2009)Bickmore, Pfeifer, and
  Jack}]{10.1145/1518701.1518891}
Timothy~W. Bickmore, Laura~M. Pfeifer, and Brian~W. Jack. 2009.
\newblock \href {https://doi.org/10.1145/1518701.1518891} {Taking the time to
  care: Empowering low health literacy hospital patients with virtual nurse
  agents}.
\newblock In \emph{Proceedings of the SIGCHI Conference on Human Factors in
  Computing Systems}, CHI '09, page 1265–1274, New York, NY, USA. Association
  for Computing Machinery.

\bibitem[{Chintagunta et~al.(2021)Chintagunta, Katariya, Amatriain, and
  Kannan}]{chintagunta2021medically}
Bharath Chintagunta, Namit Katariya, Xavier Amatriain, and Anitha Kannan. 2021.
\newblock Medically aware gpt-3 as a data generator for medical dialogue
  summarization.
\newblock In \emph{Proceedings of the Second Workshop on NLP for Medical
  Conversations}, pages 66--76.

\bibitem[{Clarke et~al.(2013)Clarke, Steege, Moore, Belden, Koopman, and
  Kim}]{clarke2013addressing}
Martina~A Clarke, Linsey~M Steege, Joi~L Moore, Jeffery~L Belden, Richelle~J
  Koopman, and Min~Soon Kim. 2013.
\newblock Addressing human computer interaction issues of electronic health
  record in clinical encounters.
\newblock In \emph{International Conference of Design, User Experience, and
  Usability}, pages 381--390. Springer.

\bibitem[{Dahlb{\"a}ck et~al.(1993)Dahlb{\"a}ck, J{\"o}nsson, and
  Ahrenberg}]{dahlback1993wizard}
Nils Dahlb{\"a}ck, Arne J{\"o}nsson, and Lars Ahrenberg. 1993.
\newblock Wizard of oz studies—why and how.
\newblock \emph{Knowledge-based systems}, 6(4):258--266.

\bibitem[{Enarvi et~al.(2020)Enarvi, Amoia, Teba, Delaney, Diehl, Hahn, Harris,
  McGrath, Pan, Pinto et~al.}]{enarvi2020generating}
Seppo Enarvi, Marilisa Amoia, Miguel Del-Agua Teba, Brian Delaney, Frank Diehl,
  Stefan Hahn, Kristina Harris, Liam McGrath, Yue Pan, Joel Pinto, et~al. 2020.
\newblock Generating medical reports from patient-doctor conversations using
  sequence-to-sequence models.
\newblock In \emph{Proceedings of the first workshop on NLP for medical
  conversations}, pages 22--30.

\bibitem[{Jalil et~al.(2019)Jalil, Myers, Atkinson, and
  Soden}]{info:doi/10.2196/humanfactors.9481}
Sakib Jalil, Trina Myers, Ian Atkinson, and Muriel Soden. 2019.
\newblock \href {https://doi.org/10.2196/humanfactors.9481} {Complementing a
  clinical trial with human-computer interaction: Patients' user experience
  with telehealth}.
\newblock \emph{JMIR Hum Factors}, 6(2):e9481.

\bibitem[{Joshi et~al.(2020)Joshi, Katariya, Amatriain, and
  Kannan}]{joshi2020dr}
Anirudh Joshi, Namit Katariya, Xavier Amatriain, and Anitha Kannan. 2020.
\newblock Dr. summarize: Global summarization of medical dialogue by exploiting
  local structures.
\newblock In \emph{Proceedings of the 2020 Conference on Empirical Methods in
  Natural Language Processing: Findings}, pages 3755--3763.

\bibitem[{Lackey et~al.(2014)Lackey, Moshiri, Pandey, Lall, Lalwani, and
  Bhargava}]{LACKEY2014481}
Amanda~E. Lackey, Mariam Moshiri, Tarun Pandey, Chandana Lall, Neeraj Lalwani,
  and Puneet Bhargava. 2014.
\newblock \href {https://doi.org/https://doi.org/10.1016/j.jacr.2013.11.020}
  {Productivity, part 1: Getting things done, using e-mail, scanners, reference
  managers, note-taking applications, and text expanders}.
\newblock \emph{Journal of the American College of Radiology}, 11(5):481--489.

\bibitem[{Lewis et~al.(2020)Lewis, Liu, Goyal, Ghazvininejad, Mohamed, Levy,
  Stoyanov, and Zettlemoyer}]{lewis-etal-2020-bart}
Mike Lewis, Yinhan Liu, Naman Goyal, Marjan Ghazvininejad, Abdelrahman Mohamed,
  Omer Levy, Veselin Stoyanov, and Luke Zettlemoyer. 2020.
\newblock \href {https://doi.org/10.18653/v1/2020.acl-main.703} {{BART}:
  Denoising sequence-to-sequence pre-training for natural language generation,
  translation, and comprehension}.
\newblock In \emph{Proceedings of the 58th Annual Meeting of the Association
  for Computational Linguistics}, pages 7871--7880, Online. Association for
  Computational Linguistics.

\bibitem[{Lin(2004)}]{lin2004rouge}
Chin-Yew Lin. 2004.
\newblock Rouge: A package for automatic evaluation of summaries.
\newblock In \emph{Text summarization branches out}, pages 74--81.

\bibitem[{MacAvaney et~al.(2019)MacAvaney, Sotudeh, Cohan, Goharian, Talati,
  and Filice}]{macavaney2019ontology}
Sean MacAvaney, Sajad Sotudeh, Arman Cohan, Nazli Goharian, Ish Talati, and
  Ross~W Filice. 2019.
\newblock Ontology-aware clinical abstractive summarization.
\newblock In \emph{Proceedings of the 42nd International ACM SIGIR Conference
  on Research and Development in Information Retrieval}, pages 1013--1016.

\bibitem[{Megges et~al.(2018)Megges, Freiesleben, Rösch, Knoll, Wessel, and
  Peters}]{MEGGES2018636}
Herlind Megges, Silka~Dawn Freiesleben, Christina Rösch, Nina Knoll, Lauri
  Wessel, and Oliver Peters. 2018.
\newblock \href {https://doi.org/https://doi.org/10.1016/j.trci.2018.10.002}
  {User experience and clinical effectiveness with two wearable global
  positioning system devices in home dementia care}.
\newblock \emph{Alzheimer's \& Dementia: Translational Research \& Clinical
  Interventions}, 4:636--644.

\bibitem[{Menachemi and Brooks(2006)}]{menachemi2006reviewing}
Nir Menachemi and Robert~G Brooks. 2006.
\newblock Reviewing the benefits and costs of electronic health records and
  associated patient safety technologies.
\newblock \emph{Journal of medical systems}, 30(3):159--168.

\bibitem[{Moramarco et~al.(2021)Moramarco, Korfiatis, Savkov, and
  Reiter}]{moramarco2021preliminary}
Francesco Moramarco, Alex~Papadopoulos Korfiatis, Aleksandar Savkov, and Ehud
  Reiter. 2021.
\newblock A preliminary study on evaluating consultation notes with
  post-editing.
\newblock In \emph{Proceedings of the Workshop on Human Evaluation of NLP
  Systems (HumEval)}, pages 62--68.

\bibitem[{Moramarco et~al.(2022)Moramarco, Papadopoulos~Korfiatis, Perera,
  Juric, Flann, Reiter, Belz, and Savkov}]{moramarco2022human}
Francesco Moramarco, Alex Papadopoulos~Korfiatis, Mark Perera, Damir Juric,
  Jack Flann, Ehud Reiter, Anya Belz, and Aleksandar Savkov. 2022.
\newblock Human evaluation and correlation with automatic metrics in
  consultation note generation.
\newblock In \emph{Proceedings of the 60th Annual Meeting of the Association
  for Computational Linguistics}.

\bibitem[{Nallapati et~al.(2016)Nallapati, Zhou, dos Santos, Gul{\c{c}}ehre,
  and Xiang}]{nallapati2016abstractive}
Ramesh Nallapati, Bowen Zhou, Cicero dos Santos, {\c{C}}a{\u{g}}lar
  Gul{\c{c}}ehre, and Bing Xiang. 2016.
\newblock Abstractive text summarization using sequence-to-sequence rnns and
  beyond.
\newblock In \emph{Proceedings of The 20th SIGNLL Conference on Computational
  Natural Language Learning}, pages 280--290.

\bibitem[{Palmieri and Stern(2009)}]{palmieri_lies_2009}
John~J. Palmieri and Theodore~A. Stern. 2009.
\newblock \href {https://doi.org/10.4088/PCC.09r00780} {Lies in the
  {Doctor}-{Patient} {Relationship}}.
\newblock \emph{Primary Care Companion to The Journal of Clinical Psychiatry},
  11(4):163--168.

\bibitem[{Papadopoulos~Korfiatis et~al.(2022)Papadopoulos~Korfiatis, Moramarco,
  Sarac, and Savkov}]{primock2022}
Alex Papadopoulos~Korfiatis, Francesco Moramarco, Radmila Sarac, and Aleksandar
  Savkov. 2022.
\newblock (in press): Primock57: A dataset of primary care mock consultations.
\newblock In \emph{Proceedings of the 60th Annual Meeting of the Association
  for Computational Linguistics}.

\bibitem[{Pearce et~al.(2016)Pearce, Ferguson, George, and
  Langford}]{pearce2016essential}
Patricia~F Pearce, Laurie~Anne Ferguson, Gwen~S George, and Cynthia~A Langford.
  2016.
\newblock The essential soap note in an ehr age.
\newblock \emph{The Nurse Practitioner}, 41(2):29--36.

\bibitem[{Portet et~al.(2009)Portet, Reiter, Gatt, Hunter, Sripada, Freer, and
  Sykes}]{portet2009automatic}
Fran{\c{c}}ois Portet, Ehud Reiter, Albert Gatt, Jim Hunter, Somayajulu
  Sripada, Yvonne Freer, and Cindy Sykes. 2009.
\newblock Automatic generation of textual summaries from neonatal intensive
  care data.
\newblock \emph{Artificial Intelligence}, 173(7-8):789--816.

\bibitem[{Reiter and Belz(2009)}]{reiter-belz-2009-investigation}
Ehud Reiter and Anya Belz. 2009.
\newblock \href {https://doi.org/10.1162/coli.2009.35.4.35405} {An
  investigation into the validity of some metrics for automatically evaluating
  natural language generation systems}.
\newblock \emph{Computational Linguistics}, 35(4):529--558.

\bibitem[{Reiter et~al.(2003)Reiter, Robertson, and Osman}]{REITER200341}
Ehud Reiter, Roma Robertson, and Liesl~M. Osman. 2003.
\newblock \href {https://doi.org/https://doi.org/10.1016/S0004-3702(02)00370-3}
  {Lessons from a failure: Generating tailored smoking cessation letters}.
\newblock \emph{Artificial Intelligence}, 144(1):41--58.

\bibitem[{Stowers and Mouloua(2018)}]{stowers2018human}
Kimberly Stowers and Mustapha Mouloua. 2018.
\newblock Human computer interaction trends in healthcare: an update.
\newblock In \emph{Proceedings of the International Symposium on Human Factors
  and Ergonomics in Health Care}, volume~7, pages 88--91. SAGE Publications
  Sage CA: Los Angeles, CA.

\bibitem[{Yim and Yetisgen-Yildiz(2021)}]{yim2021towards}
Wen-wai Yim and Meliha Yetisgen-Yildiz. 2021.
\newblock Towards automating medical scribing: Clinic visit dialogue2note
  sentence alignment and snippet summarization.
\newblock In \emph{Proceedings of the Second Workshop on NLP for Medical
  Conversations}, pages 10--20.

\bibitem[{Zhang et~al.(2021)Zhang, Negrinho, Ghosh, Jagannathan, Hassanzadeh,
  Schaaf, and Gormley}]{zhang2021leveraging}
Longxiang Zhang, Renato Negrinho, Arindam Ghosh, Vasudevan Jagannathan,
  Hamid~Reza Hassanzadeh, Thomas Schaaf, and Matthew~R Gormley. 2021.
\newblock Leveraging pretrained models for automatic summarization of
  doctor-patient conversations.
\newblock In \emph{Findings of the Association for Computational Linguistics:
  EMNLP 2021}, pages 3693--3712.

\bibitem[{Zhang et~al.(2018)Zhang, Ding, Qian, Manning, and
  Langlotz}]{zhang2018learning}
Yuhao Zhang, Daisy~Yi Ding, Tianpei Qian, Christopher~D Manning, and Curtis~P
  Langlotz. 2018.
\newblock Learning to summarize radiology findings.
\newblock In \emph{Proceedings of the Ninth International Workshop on Health
  Text Mining and Information Analysis}, pages 204--213.

\end{thebibliography}
\bibliographystyle{acl_natbib}

\appendix
\onecolumn
\section{Appendix}
\label{sec:appendix}

\subsection{Patient consent form}
\label{sec:consent-form}

The following is the consent form we obtained from patients prior to a consultation observation. 

\emph{``At Babylon, we sometimes have training doctors and support staff observe our clinics for learning how we use our internal systems. I have [X] participants observing today who have completed all relevant security checks and received specific training to enable them to safely observe consultations. So before we continue, we always check consent with patients - are you happy to proceed?''}

\subsection{Example mock transcript and corresponding note}
\label{app:transcript-note}

\renewcommand{\cellalign}{l}
\begin{table*}[h]
    \setlength{\tabcolsep}{4pt} 
    \def\arraystretch{1.3}
    \centering
    \begin{tabular}{l|p{8.5cm}|p{5cm}}
        \multicolumn{2}{c|}{\cellcolor{blue!25}\textbf{Transcript}} & \multicolumn{1}{c}{\cellcolor{blue!25}\textbf{Note}} \\\hline
          \cellcolor{blue!5}Clinician & \cellcolor{blue!5}So, um, tell me what's been going on. You've been saying there's a problem with your hearing. Is that right? & \multirow{8}{*}{\makecell{History:\\
          Hx of difficulty hearing left ear\\
          for 6 weeks with tinnitus and\\
          slight nausea/ dizziness.\\
          One previous similar episode in\\ the past- resolved spontaneously.\\
          No discharge/fever/itchiness/pain\\
          Doesn't use cotton wool buds\\
          No Pmhx of note\\
          Ex: Looks well, not in pain.\\
          \textcolor{violet}{Imp: need to exclude impacted}\\ \textcolor{violet}{wax in ear canal first}\\
          \textcolor{violet}{Pln: for face to face GP}\\ \textcolor{violet}{appointment in 5 days to examine}\\
          \textcolor{violet}{ear}\\
          \textcolor{violet}{If any problems in interim to}\\ \textcolor{violet}{ring us back}\\
          \textcolor{violet}{Pt happy with and understands}\\\textcolor{violet}{plan}}}\\

          \cline{1-2}
         
         Patient & Yeah, so I just feel I can't really hear as well as I used to, like my hearing is kind of deteriorating in some way. \\\cline{1-2}
         \cellcolor{blue!5}Clinician & \cellcolor{blue!5}Right, OK. How long has this been going on for? \\\cline{1-2}
         Patient & Uh about six weeks. \\\cline{1-2}
         \cellcolor{blue!5}Clinician & \cellcolor{blue!5}Six weeks, OK. Um, and before that have you had any hearing problem at all? \\\cline{1-2}
         Patient & Um I had something maybe, about a year ago, but it only lasted a couple of days, it wasn't anything as long as this.\\\cline{1-2}
         \cellcolor{blue!5}Clinician & \cellcolor{blue!5}Right, OK, OK. And, um, in this six week period, have you had anything else happen? Have you had any other ear symptoms at all? \\\cline{1-2}
         Patient & Um, I occasionally get like a ringing in my left ear, uh just on the one side and um there's actually been a few times when I felt kind of a bit sick or a bit dizzy as well.\\
    \end{tabular}
    \caption{Snippet of a mock consultation transcript and the corresponding note, written by the consulting clinician. Text in violet demarks the diagnosis and treatment advice, which we exclude in the Live Test.}
    \label{tab:transcript-note}
\end{table*}

\begin{figure}[ht]
\begin{subfigure}{\linewidth}
  \centering
  \caption{Design 1: Live transcript of the consultation (red box) and generated note displayed at the end (green box).}
  \includegraphics[width=.8\linewidth]{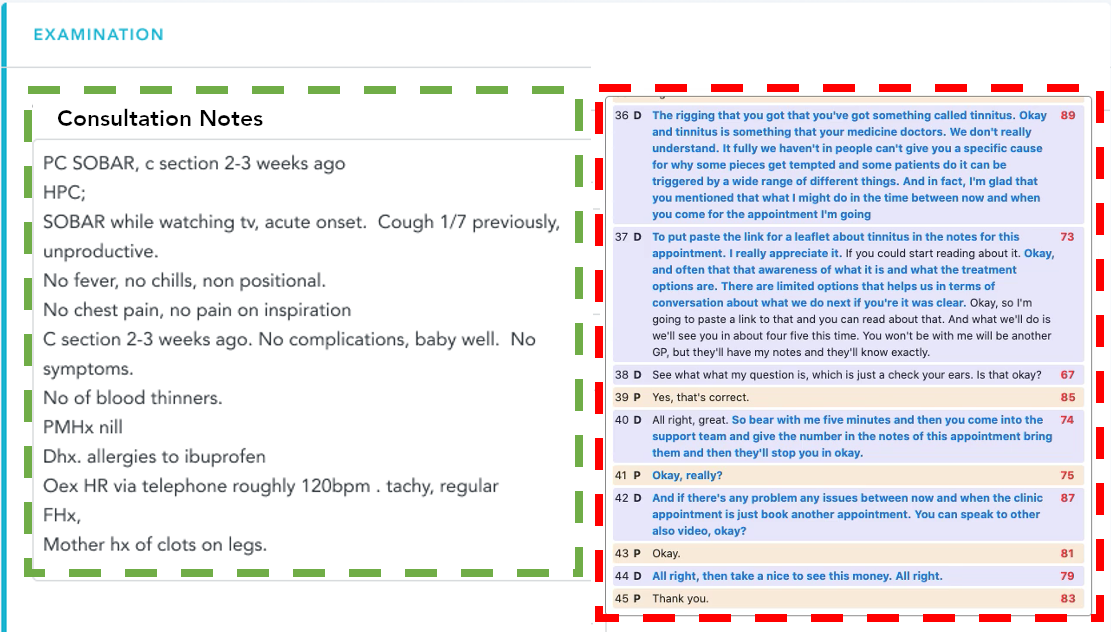}  
  \label{fig:sub-first}
\end{subfigure}
\begin{subfigure}{1\linewidth}
  \centering
  \caption{Design 2: No feedback during the consultation and generated note displayed at the end (green box).}
  \includegraphics[width=.8\linewidth]{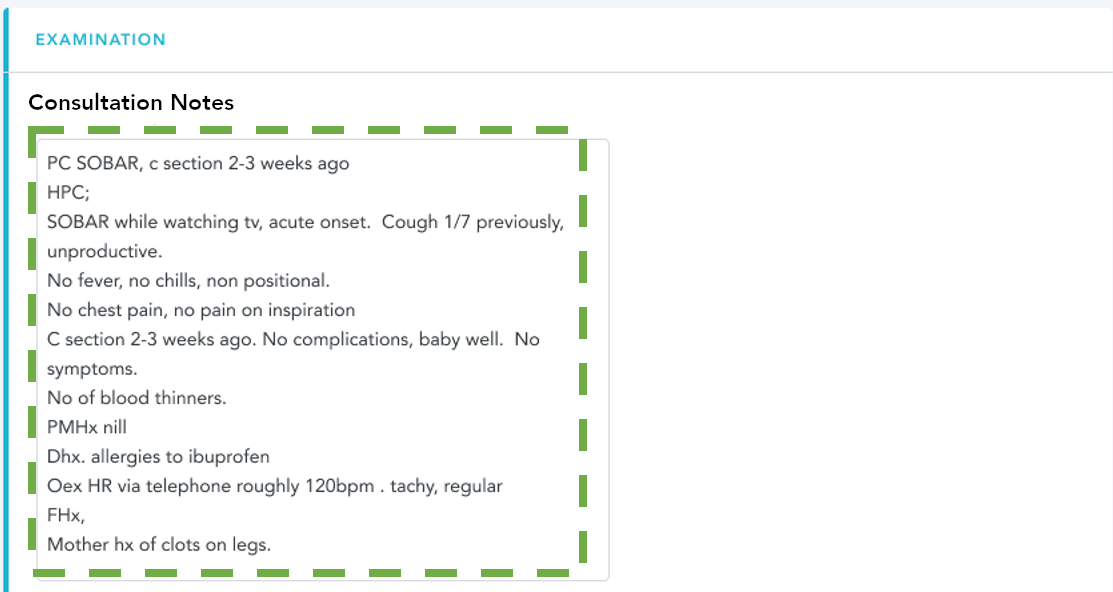}  
  \label{fig:sub-second}
\end{subfigure}
\begin{subfigure}{1\linewidth}
  \centering
  \caption{Design 3: Live note generated during the consultation (red box).}
  \includegraphics[width=.8\linewidth]{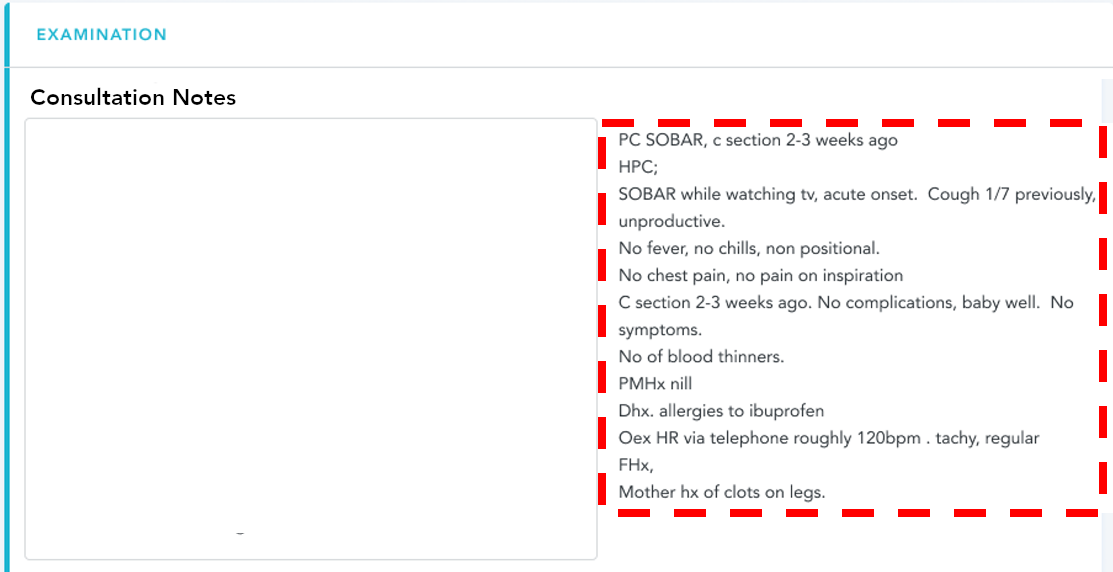}  
  \label{fig:sub-third}
\end{subfigure}
\caption{The three designs presented to clinicians in Discovery Phase 2. A red box indicates a textual pane updated during the consultation. A green box indicates a pane updated at the end of the consultation.}
\label{fig:ui-designs}
\end{figure}

\end{document}